\documentclass[10pt,conference]{IEEEtran}
\IEEEoverridecommandlockouts
\usepackage{cite}
\usepackage{amsmath,amssymb,amsfonts}
\usepackage{algorithmic}
\usepackage{graphicx}
\usepackage{textcomp}
\usepackage{xcolor}

\usepackage{engord}
\usepackage{url}

\def\BibTeX{{\rm B\kern-.05em{\sc i\kern-.025em b}\kern-.08em
    T\kern-.1667em\lower.7ex\hbox{E}\kern-.125emX}}
\begin{document}

\title{Data Analytics and Machine Learning Methods, Techniques and Tool for Model-Driven Engineering of Smart IoT Services}

\author{\IEEEauthorblockN{Armin Moin}
\IEEEauthorblockA{\textit{Department of Informatics, Technical University of Munich (TUM), Germany} \\
moin@in.tum.de}
}

\maketitle

\begin{abstract}
This doctoral dissertation proposes a novel approach to enhance the development of smart services for the Internet of Things (IoT) and smart Cyber-Physical Systems (CPS). The proposed approach offers abstraction and automation to the software engineering processes, as well as the Data Analytics (DA) and Machine Learning (ML) practices. This is realized in an integrated and seamless manner. We implement and validate the proposed approach by extending an open source modeling tool, called ThingML. ThingML is a domain-specific language and modeling tool with code generation for the IoT/CPS domain. Neither ThingML nor any other IoT/CPS modeling tool supports DA/ML at the modeling level. Therefore, as the primary contribution of the doctoral dissertation, we add the necessary syntax and semantics concerning DA/ML methods and techniques to the modeling language of ThingML. Moreover, we support the APIs of several ML libraries and frameworks for the automated generation of the source code of the target software in Python and Java. Our approach enables platform-independent, as well as platform-specific models. Further, we assist in carrying out semi-automated DA/ML tasks by offering Automated ML (AutoML), in the background (in expert mode), and through model-checking constraints and hints at design-time. Finally, we consider three use case scenarios from the domains of network security, smart energy systems and energy exchange markets.
\end{abstract}


\section{Introduction}\label{introduction}
Developing software for the \textit{Internet of Things (IoT)} and \textit{smart Cyber-Physical Systems (CPS)} is a relatively complex task. There exist various reasons for this complexity. First, the distributed nature of such systems leads to many challenges. The key concerns and challenges associated with almost any distributed computing system, particularly network-centric service-oriented architectures, include heterogeneity (e.g., with respect to the hardware architectures, operating systems, communication protocols, programming languages and APIs), openness, security and trustworthiness, scalability, failure handling and fault tolerance, concurrency, transparency, quality of service (including availability, reliability, performance and adaptability), as well as privacy \cite{Coulouris+2011}. Second, smart CPS, which merge the physical and the virtual worlds using Information and Communication Technologies (ICT) in smart ways \cite{GeisbergerBroy2014}, pose their own unique challenges, e.g., concerning their \textit{cross-dimension} and cascade effects (i.e., cross-domain, cross-technology, cross-organization and cross-functional), their \textit{live-dimension} (i.e., re-configuration, re-deployment, live-update, live-enhancement and live-extension), and their \textit{self-dimension}, referred to as \textit{Self-* capabilities} (i.e., self-documenting, self-monitoring/diagnosis, self-optimizing, self-healing and self-adapting/training) \cite{Schaetz2014}. In addition, the Internet is continuously expanding into a pervasive, global and collaborative network of almost every entity/actor (i.e., object or \textit{thing}) surrounding us, including a huge number of sensors and actuators in our smart environments, thus incorporating and integrating them into the IoT. Computing devices networked into the IoT are even much more heterogeneous, e.g., their power supplies/storages, processors, main memories and data storage capacities can have a very broad spectrum. Thus, unprecedented challenges are faced by the designers and developers of such systems \cite{Atzori+2010}. 

Prior research \cite{Schaetz2014, Harrand+2016} has proven that \textit{Model-Driven Engineering (MDE)} can be of significant benefit in the domain of IoT/CPS. MDE can hide complexity through abstraction, support communication among stakeholders, and assist with design, early validation/verification/testing, simulation, implementation, and maintenance/evolution in an automated or semi-automated manner. This has motivated \textit{Model-Driven Software Engineering (MDSE)}, particularly \textit{Domain-Specific Modeling (DSM)} with full code generation \cite{KellyTolvanen2008} for the IoT/CPS, as the focus of this PhD study. Moreover, we adopt three use case domains for the creation of our solution, as well as the demonstration of the results and the validation of the proposed approach: network security, smart energy systems and energy exchange markets.

One state-of-the-art domain-specific modeling tool for the IoT/CPS has been delivered by the open source project, \textit{ThingML (Things Modeling Language)} \cite{ThingML, Harrand+2016}. This doctoral dissertation has targeted the problem of lack of support for Data Analytics (DA) and Machine Learning (ML) at the modeling level of ThingML and other IoT/CPS modeling tools. Given the crucial role of ML in Artificial Intelligence (AI), and the prominent effect of AI on IoT/CPS, this shortcoming in the state of the art represents a critical technological gap that needs to be addressed. Therefore, this doctoral dissertation extends ThingML with DA/ML methods and techniques to facilitate the development of smart IoT services and CPS applications \cite{Moin+2018,Moin+2020}. Concretely, we propose a methodology, a modeling language with extended syntax and semantics, as well as a number of code generators, based on ThingML.

The rest of this doctoral symposium paper is structured as follows: Section \ref{hypotheses} states our main hypotheses. Moreover, we review the state of the art and mention our expected contributions in Section \ref{related-work-and-contributions}. Further, Section \ref{evaluation-plan} illustrates our plan for the validation and evaluation methodology. Additionally, a brief description of the results achieved so far is provided in Section \ref{preliminary-results}. Finally, we conclude and propose the planned timeline for the completion of the doctoral study in Section~\ref{completion-plan}.

\section{Hypotheses}\label{hypotheses}

We aim to explore \textit{models} in DA/ML and \textit{models} in software engineering, particularly MDSE, in order to find synergies and validate the research hypotheses. This doctoral dissertation has three main hypotheses:

\paragraph*{\engordnumber{1} Hypothesis}DA/ML concepts can be added to the modeling level of an existing state-of-the-art modeling language and tool for the domain of IoT/CPS, e.g., ThingML, while still supporting the automated generation of the full source code of the target services/applications. Hence, software models in MDSE can be employed to generate ML models and/or deal with them.

\paragraph*{\engordnumber{2} Hypothesis}
The option of keeping two separate modeling layers, namely Platform-Independent Model (PIM) and Platform-Specific Model (PSM) \cite{OMGMDA2014} is feasible and deployable. Specifically, for the platform choices concerning DA/ML, we can support both Java (e.g., via the APIs of WEKA \cite{Hall+2009} / MOA \cite{Bifet+2010}), and Python (e.g., via the APIs of Scikit-Learn \cite{Pedregosa+2011} and Tensorflow \cite{Abadi+2015} / Keras \cite{Chollet+2015} / Pytorch \cite{Paszke+2017}) for code generation. 

\paragraph*{\engordnumber{3} Hypothesis}
The workflows and pipelines regarding the DA/ML practices of the users of the resulting tool for the said use case domains can be partially automated. In particular, we can offer Automated ML (AutoML) to assist in the data preparation tasks of time series data, as well as in the choice of ML models/algorithms and hyper-parameter optimization. Also, we can provide hints and live feedback regarding DA/ML practices at design-time via the model-checking constraints, exception handling mechanisms and the model editors.

\section{Related Work and Contributions}\label{related-work-and-contributions}
\paragraph*{Computer-Aided Software Engineering (CASE)}
The most relevant work in this regard has been delivered via the open source project \textit{ThingML} \cite{ThingML, Harrand+2016}. Various research projects, including \textit{HEADS} \cite{HEADS} have contributed to ThingML over the past years. Using the Domain-Specific Language (DSL) of ThingML, one may specify the so-called Heterogeneous and Distributed (HD) services, i.e., IoT services on a higher level of abstraction. Once the model is complete and valid, the code generation framework can produce the full implementation of the target software in a number of programming languages, e.g., Java, Javascript and C, for a range of supported target platforms, e.g., Linux, Arduino and Android, and communication protocols, e.g., HTTP, CoAP and MQTT. The generated code includes the configuration scripts and is ready to be deployed. As mentioned in Section \ref{introduction}, ThingML and other IoT/CPS modeling tools come short of supporting DA/ML practitioners at the modeling level, although DA and ML are quite vital for IoT/CPS.

\paragraph*{DA/ML}
Increasing the level of abstraction by offering higher level APIs has also been applied to the fields of DA and ML. For instance, there exist a variety of open source libraries and frameworks for ML, such as \textit{Scikit Learn} \cite{Pedregosa+2011}, \textit{TensorFlow} \cite{Abadi+2015}, \textit{Theano} \cite{Theano2016}, \textit{Pytorch} \cite{Paszke+2017}, \textit{WEKA} \cite{Hall+2009}, \textit{MOA} \cite{Bifet+2010}, \textit{Mahout} \cite{Mahout} and \textit{MLlib} \cite{Meng+2016}. They offer ML experts a higher level of abstraction. Furthermore, \textit{Keras} \cite{Chollet+2015} and \textit{MXNet} \cite{Chen+2015} provide yet another higher level of abstraction to practitioners by supporting the APIs of several frameworks and libraries. For instance, Keras supports both TensorFlow and Theano as its backend. Besides, we observed the emergence of a number of workflow designers and tools for partial automation of data analytics pipelines over the past two decades. The \textit{Konstanz Information Miner (KNIME)} \cite{Berthold+2009} is one of the best open-source examples. Note that code generation is also supported to some extent. However, it is not comparable to the systematic approach of the MDSE paradigm. Most importantly, the said workflow designers do not cover any aspect of software beyond DA/ML in their modeling. Hence, they are not sufficient for the IoT/CPS domain.

\paragraph*{Model-Based ML}
Bishop \cite{Bishop2013} proposed the idea of \textit{model-based ML}, thus trying to boost ML models to take over the role of software models in MDSE, e.g., by generating the full source code of the target application out of them. That approach has been implemented in the open source framework \textit{Infer.NET} \cite{InferNet, Bishop2013}, which supported the probabilistic programming paradigm. The user of Infer.NET should specify a Probabilistic Graphical Model (PGM) via the provided DSL. The framework could then generate the source code of the target software application out of the PGM. However, we argue that this approach is highly restrictive, since PGMs are not expressive enough to let the user specify smart IoT services / CPS applications sufficiently. Hence, our proposed approach is vice versa, i.e., we enhance software models to become capable of producing and dealing with ML models. Further, Infer.NET is inflexible, both in terms of the choice of the ML models and algorithms (limited to PGMs, while other ML models, such as neural networks are currently widely used in industry), and with respect to the code generation (i.e., the generated code can only be in C\#). Last but not least, Breuker \cite{Breuker2014} conducted an exploratory analysis of DSLs for ML and proposed an initial conceptualization of a DSL, based on Infer.NET.

\paragraph*{Expected Contributions}
We propose a novel approach, called \textit{ThingML+}, which addresses the above-mentioned shortcomings of the related work (see Table \ref{tab:1}). In particular, we adopt the idea of model-based ML \cite{Bishop2013}, but, as stated above, in a very different way. Instead of using ML models as software models, we enhance software models to become capable of supporting ML, e.g., by generating the appropriate ML models for the task at hand, and training them automatically using existing and/or incoming data. In addition, the user may bring his or her own trained ML model, e.g., an Artificial Neural Network (ANN), which is already trained on a dataset, and simply connect that to the software model. This is called a mixed MDSE/non-MDSE approach, since the software model may not include all the necessary details to be able to generate the ANN and the source code that is necessary to train it on existing or new data instances in the future. Finally, we implement our proposed approach in an open source tool, called \textit{ML-Quadrat} \cite{ML-Quadrat}. Similar to ThingML, ML-Quadrat is also built on top of the Eclipse Modeling Framework (EMF). From the research perspective, we validate the three hypotheses mentioned in Section \ref{hypotheses} and evaluate the proposed approach.

\begin{table}[htbp]
\caption{Comparison of the state of the art with the proposed approach (ThingML+)}
\begin{center}
\begin{tabular}{|p{2cm}|p{2cm}|p{2cm}|p{1cm}|}
	\hline	
	& CASE / MDSE w. full-code gen. & DA / ML w/o AutoML & IoT / CPS \\
	\hline
	ThingML & \checkmark &  & \checkmark \\
	\hline
	KNIME, WEKA, TensorFlow, etc. & & \checkmark & \\
	\hline
	INFER.NET & \checkmark & \checkmark & \\
	\hline
	ThingML+ & \checkmark & \checkmark & \checkmark \\
	\hline
\end{tabular}
\label{tab:1}
\end{center}
\end{table}

\section{Validation and Evaluation Plan}\label{evaluation-plan}
Newman \cite{Newman1994} highlighted four broad categories for engineering research: (i) Enhanced Analytical Modeling Techniques (abbreviated as EM), (ii) Enhanced Solutions (abbreviated as ES), (iii) Radical Solutions (abbreviated as RS), and (iv) Enhanced Tools and Methods (abbreviated as ET). Although his work originated from the field of Human Computer Interaction (HCI), it was not specific to HCI, but rather for any engineering research work. The three hypotheses mentioned in Section \ref{hypotheses} indicate that our research is primarily about the latter class, i.e., enhancement of tools and methods. Hence, as suggested by Newman \cite{Newman1994}, we shall validate the proposed approach by implementing our modeling tool and methods, thus showing the feasibility, e.g., for the selected use case domains, through a number of working examples. Note that we employ the use case domains of network security and smart energy systems for creating our solution (see Section \ref{preliminary-results}). However, for the validation and evaluation, we use the use case domains of smart energy systems (with different datasets) and energy stock markets.

Furthermore, we shall validate the usability of the proposed approach. This will be done in two ways. First, we shall demonstrate that the automatically generated code performs on average better or at least as good as the state of the art (i.e., manually developed software). In fact, we expect our approach to outperform the state of the art, since the proposed AutoML functionalities shall lead to performance leaps, e.g., due to the optimized data preparation efforts, as well as model/algorithm and hyper-parameter choices for ML. Second, we shall also study the user experience and confirm that the intended audience, who are mostly software developers without deep knowledge and skills in DA/ML or specific target IoT platforms, are more satisfied with using our proposed solution than traditional approaches, i.e., manual software development. For the validation of the usability and user experience, we shall conduct an empirical study with human participants.

Therefore, we shall validate each of the said hypotheses as follows: (i) For the first hypothesis, we shall show that the source code that will be generated out of each model instance does not require any further manual changes, and is ready to be deployed and executed on the respective target platforms. (ii) Concerning the second hypothesis, we shall illustrate that model instances may be Platform-Independent Models (PIM) or Platform-Specific Models (PSM) via examples. Usually, we let the user of our tool import a PIM into a PSM and augment it with the platform-specific details to make it a complete model instance for code generation. (iii) Regarding the third hypothesis, we shall compare the performance and efficiency of the automatically generated code in the AutoML mode with the manually developed code. In particular, we shall compare the accuracy, precision, recall and F1-measure in both cases and confirm that the AutoML functionalities improve those metrics. 

Finally, GQM (Goal, Question, Metric) \cite{Basili+1994} provided a measurement model on the three levels of conceptual (Goal), operational (Question) and quantitative (Metric). We shall consider this approach in order to devise our concrete validation plan.

\section{Preliminary Results}\label{preliminary-results}
So far, we have published two peer-reviewed papers for this dissertation. The first one, \textit{ThingML+} (2018) \cite{Moin+2018} was a position paper (3 pages), which illustrated the core concept. Moreover, the second one, \textit{ThingML2} (2020) \cite{Moin+2020} was a poster together with a companion extended abstract (2 pages), which demonstrated the proof of concept and the validation of the first hypothesis. In the following, we present the current status of our ongoing research and development. Our research prototype is available as open source software with Apache License 2.0. Please check out our repository on Github \cite{ML-Quadrat}.


\subsection*{Use Cases}
We focus on three use case domains as follows. The main use case domain is smart energy systems, which will be used for both creating our solution and validating it, but with different datasets. However, the network security use case shall be additionally employed for creating the solution, whereas the energy exchange markets use case shall further validate the proposed approach. 

\paragraph*{Network Security}
Distributed Denial-of-Service (DDOS) attacks are among the most prevalent types of malicious cyber activities to prevent legitimate access to Internet-based services. We consider a simple client-server interaction, where the server simply waits for ping messages from clients and responds to each ping message with a pong message. We enhance this ping-pong example using ML. Concretely, we empower the server with the ability to check whether the client is likely to be an attacker. If this is the case, the server will simply ignore the ping message of the client. Otherwise, the server may send out the pong message in response to the client. The ML model, which is trained using an existing dataset of IP addresses and timestamps of previously received ping messages over a time period (e.g., several months), uses the IP address and the timestamp of the ping messages to make predictions. It predicts a boolean value for each ping message. If the sender (client) is more likely to be an attacker, the predicted boolean value will be true. Otherwise, it will be false. The pre-existing dataset must be sufficiently long to expect meaningful predictions by the ML model. This use case serves for illustration purpose and is not a real-world use case scenario (unlike the other two use cases). 

\begin{figure}[htbp]
	\centerline{\includegraphics[width=0.4\textwidth]{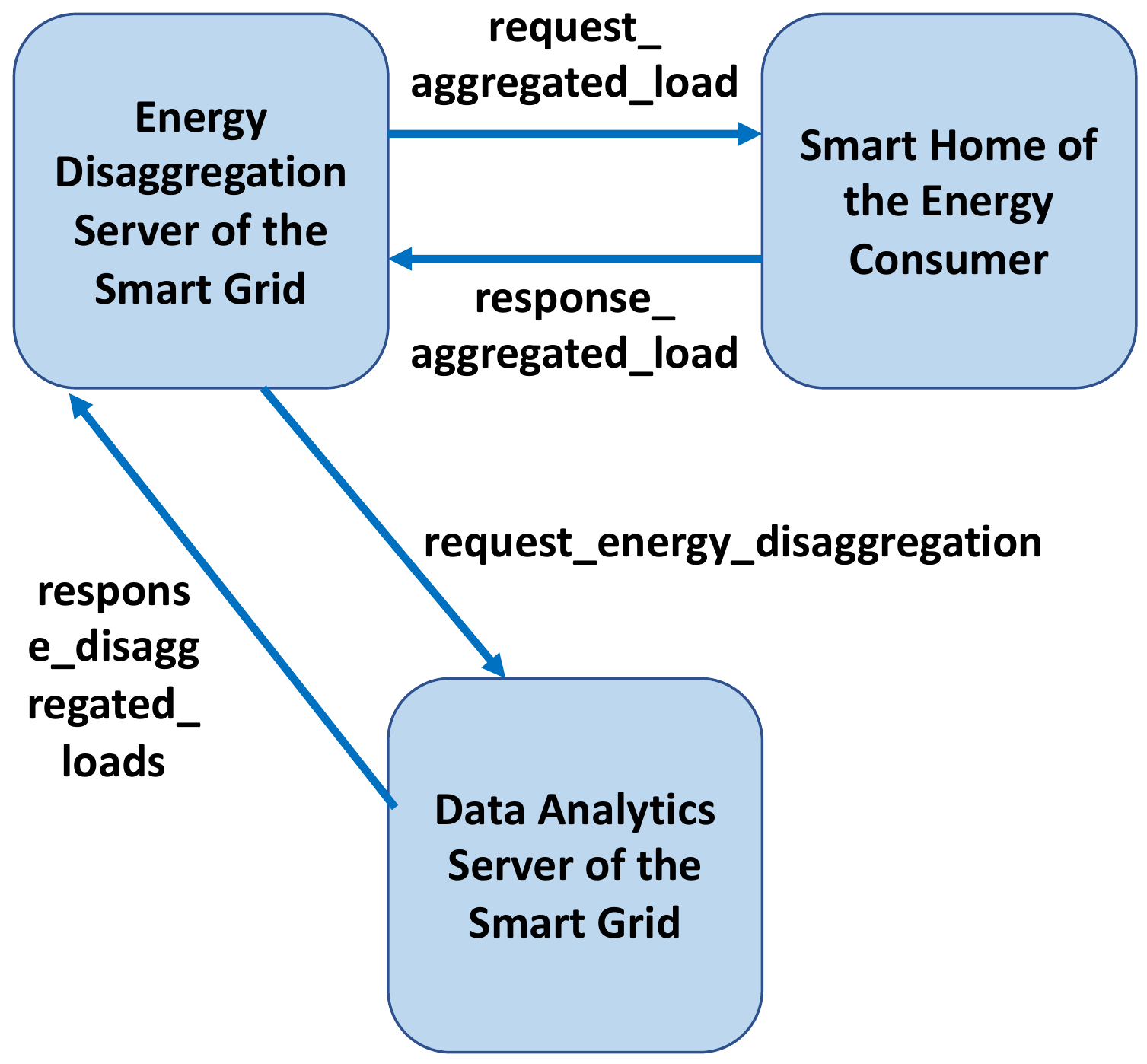}}
	\caption{The architecture of our example for NIALM.}
	\label{fig:NIALM_Things}
\end{figure}

\begin{figure}[htbp]
	\centerline{\includegraphics[width=0.4\textwidth]{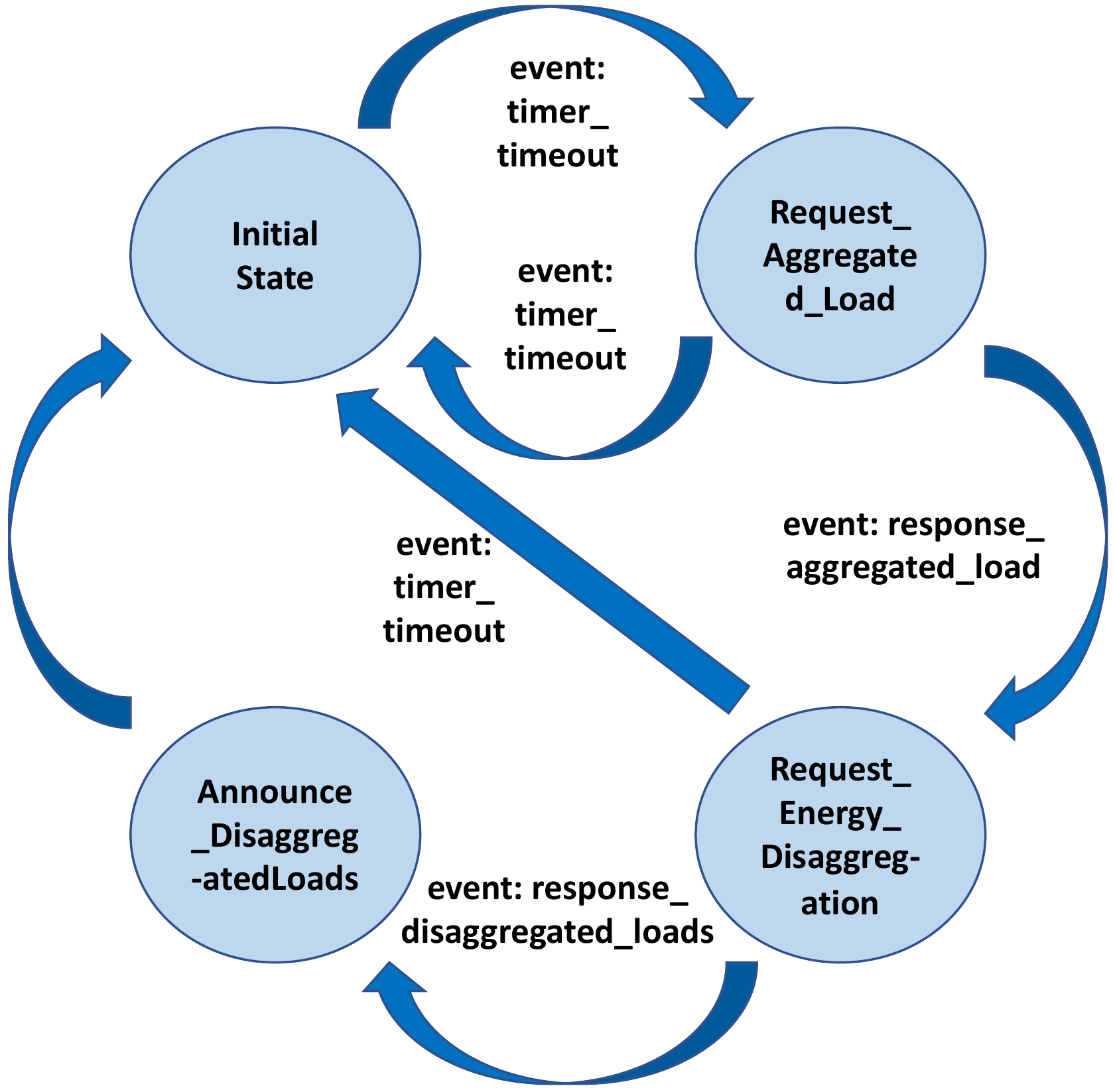}}
	\caption{The behavioral model of the energy disaggregation server.}
	\label{fig:NIALM_SmartGrid}
\end{figure}

\paragraph*{Smart Energy Systems}
We are interested in the problem of \textit{Non-Intrusive (Appliance) Load Monitoring (NIALM/NILM)}, also known as \textit{energy disaggregation}, in the context of \textit{smart grids}. Smart grids hold as an example for smart CPS, where the energy producers, energy consumers and the so-called \textit{prosumers} (i.e., those participants, who may play both roles) can be connected via the IoT, in order to optimize the overall demand and supply in the network, and make the entire grid more robust and reliable. In many cases, it is necessary or at least advantageous for the system to find out which appliance is active at each household or building that is connected to the smart grid at each point in time, and how much energy it consumes. For instance, if the system is supposed to consider certain discounts and incentives for those consumers, who refrain from using their \textit{energy-hungry} appliances, e.g., air conditioners or cryptocurrency mining servers during the \textit{peak demand} times of the grid, and have agreed to use certain appliances, e.g., dish washers, only in the \textit{off-peak} times, e.g., mid-nights, it must be possible for the system to verify whether they have indeed respected the agreements or whether any possible violations have occurred. Measuring the exact power consumption of an individual appliance is only possible, if one attaches a sensor/counter, known as a \textit{smart meter} to it. That is called the \textit{intrusive} approach, since new connected devices from the outside must be added to the household. However, it is not only costly for the owner, but can also lead to inconveniences, as well as security and privacy concerns of the occupants. In contrast, the \textit{non-intrusive} approach employs algorithms and methods to disaggregate the total power consumption signal into individual loads, thus \textit{virtually} sensing/counting the estimated power consumption of each appliance at each time. Some energy disaggregation approaches may also predict the types of appliances, which reside in each household, by comparing their power consumption patterns with certain \textit{signatures} of common home/office appliances. We use three public datasets for NIALM. The first one is the REDD dataset \cite{KolterJohnson2011} from the USA, whereas the second and the third ones are the UK-DALE \cite{KellyKnottenbelt2015} and the REFIT \cite{Murray+2017} datasets from the UK. Fig. \ref{fig:NIALM_Things} depicts the architecture of our example for NIALM. This example involves three \textit{things}: the energy disaggregation server of the smart grid, the DA/ML server of the smart grid, and the smart home of the energy consumer. The energy disaggregation server delegates the actual task of performing NIALM to the DA/ML server. Further, we illustrate the state machine model of the behavior of the energy disaggregation server in Fig. \ref{fig:NIALM_SmartGrid}. The state machines concerning the behavioral models of the other two \textit{things} are not shown here. Finally, the datasets that are used for creating the solution and validating the proposed approach are mutually exclusive.

\paragraph*{Energy Exchange Markets}
We use the open data from energy exchange markets, e.g., the European Energy Exchange (EEX) to create a prediction service for the electricity prices in the near future. This use case domain shall only be used for the validation of the proposed approach. The ML model employed here will be a sequence-to-sequence model, trained using the historic data of the dynamic electricity prices over a period of time. Given a list of current electricity prices in the market, the prediction service shall forecast the electricity prices in the near future, e.g., for the next day.  

\subsection*{Our Research Methodology}
First, we adopt the modeling language and tool provided by the ThingML project. Second, we manually develop a number of examples for the target IoT services. To this aim, we concentrate on the network security and the smart energy systems use cases. Based on that, we extract the required concepts that need to be added to the existing modeling language and tool of ThingML. Furthermore, we explore relevant standards, such as the Predictive Model Markup Language (PMML), as well as the open source libraries and frameworks for ML, specifically Scikit-Learn, Keras, TensorFlow and WEKA to select the additional concepts and platform-specific APIs that shall be added to the meta-model and the tool. Moreover, we intend to employ additional principles and guidelines in the modeling community, e.g., \cite{Selic2007} and \cite{Lagarde+2007}, to ensure that we have a proper and complete DSL.

\subsection*{Software Development Methodology at Design-Time}
A distributed system in ThingML is modeled as a set of components (i.e., \textit{things}), which communicate with each other using asynchronous message passing. In order to design a new IoT service / CPS application using ThingML, one shall specify the following: (i) The ports for message passing, the messages together with their parameters, and the properties (i.e., variables) for each \textit{thing}. (ii) The UML-like state machine (state diagram or statechart), which models the behavior of each \textit{thing}. ThingML offers an imperative platform-independent action language to support the event-driven programming paradigm on the state machines. (iii) The configuration for the entire service/application, which comprises the instances of the defined \textit{things} and the desired connections among their ports. We extend the methodology of ThingML to support DA/ML. Concretely, we let the user of our tool dedicate certain \textit{things} or \textit{thing fragments} (i.e., sub-components of \textit{things}) to DA/ML tasks. Therefore, for such \textit{things}, the user shall also specify an additional block concerning DA/ML, before the state machine that specifies the behavior of the \textit{thing}. Moreover, \textit{things} dedicated to DA/ML typically have the following states: (i) data preprocessing, (ii) training, (iii) ready for prediction, (iv) predict. If the user decides to bring his or her own trained ML model, states (i) and (ii) may be skipped.

\subsection*{Domain-Specific Modeling Language}
\begin{figure}[h!]
	\includegraphics[width=0.50\textwidth]{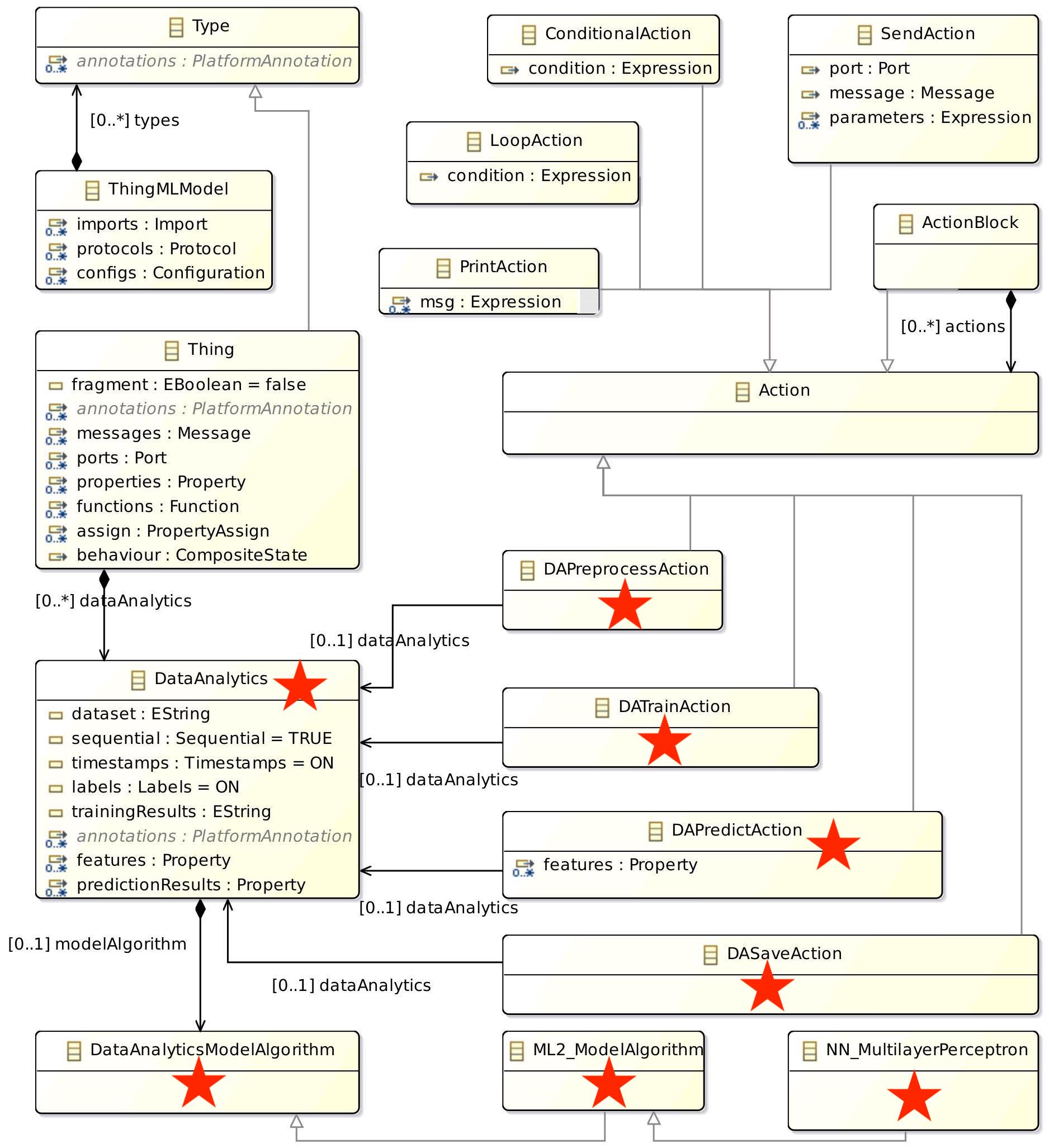}
	\caption{Part of the abstract syntax of the DSL. The red stars mark some of our extensions of the ThingML meta-model \cite{Moin+2020}.}
	\label{fig:ML2_Class_Diagram_Part}
\end{figure}

We extend the DSL of ThingML with the following items: (a) An optional DA/ML block for each \textit{thing}, which can, e.g., include the list of properties (variables), which shall be considered as the ML features/attributes, and the choice of ML model and algorithm in the expert mode. If the user is not an expert in ML, the mode shall be set to AutoML instead of the expert mode. (b) New options in the imperative action language that may be employed for event-driven programming. The new options include \textit{DA save action}, \textit{DA preprocess action}, \textit{DA train action} and \textit{DA predict action}. They can be used in the conditions for the transitions of the state machines of \textit{things} responsible for DA/ML. Fig. \ref{fig:ML2_Class_Diagram_Part} demonstrates part of the abstract syntax of our DSL \cite{Moin+2020}.

\subsection*{Code Generators}
Currently, we support code generation in Python for ML, using the APIs of Scikit-Learn and Keras (with the Tensorflow backend). The generated Python code is seamlessly integrated with the Java code using the \textit{Process Builder} API of Java. Moreover, the generated code can be easily installed and deployed using Apache Maven. All the required configuration scripts are generated out of the software model instance in an automated manner.

\subsection*{Model Editors}
We support both the customized Xtext-based textual model editor of ThingML, as well as an EMF-based graphical tree model editor. Furthermore, a web-based model editor is in development. A preliminary textual version of the web-based editor is already available.

\section{Completion Plan}\label{completion-plan}

Until the Doctoral Symposium, we will finish the implementation and validation. After the event, we will have at least 9 months for applying the feedback that is expected to be received at the event, writing the dissertation and preparing for any possible resubmission of the planned research papers. Currently, the following peer-reviewed research papers (6+ pages) are planned for the completion of the doctoral study: (i) Validation of the first hypothesis (without the empirical study); (ii) Validation of the meta-model, the second hypothesis and support for a standard Model Interchange Format (MIF), such as PMML, PFA (Portable Format for Analytics) or ONNX (Open Neural Network Exchange) for the platform-independent layer; (iii)Validation of the third hypothesis (without the empirical study); (iv) The empirical study using human participants concerning the user experience; (v) Survey and benchmarking of the state of the art for the use case domain of smart energy systems, specifically NIALM (optional journal article).

%

\section*{Acknowledgments}
The advisor, second advisor and mentor of this work are Stephan G{\"u}nnemann, Atta Badii and Marouane Sayih, respectively. The author would like to sincerely thank them for their kind supports. This work is partially funded by the German Federal Ministry of Education and Research (BMBF) through the Software Campus program (project ML-Quadrat). The author would like to thank Stephan R{\"o}ssler and his colleagues at Software AG for their help.

\bibliography{refs}

\begin{thebibliography}{10}
\providecommand{\url}[1]{#1}
\csname url@samestyle\endcsname
\providecommand{\newblock}{\relax}
\providecommand{\bibinfo}[2]{#2}
\providecommand{\BIBentrySTDinterwordspacing}{\spaceskip=0pt\relax}
\providecommand{\BIBentryALTinterwordstretchfactor}{4}
\providecommand{\BIBentryALTinterwordspacing}{\spaceskip=\fontdimen2\font plus
\BIBentryALTinterwordstretchfactor\fontdimen3\font minus
  \fontdimen4\font\relax}
\providecommand{\BIBforeignlanguage}[2]{{%
\expandafter\ifx\csname l@#1\endcsname\relax
\typeout{** WARNING: IEEEtran.bst: No hyphenation pattern has been}%
\typeout{** loaded for the language `#1'. Using the pattern for}%
\typeout{** the default language instead.}%
\else
\language=\csname l@#1\endcsname
\fi
#2}}
\providecommand{\BIBdecl}{\relax}
\BIBdecl

\bibitem{Coulouris+2011}
G.~Coulouris, J.~Dollimore, T.~Kindberg, and G.~Blair, \emph{Distributed
  Systems: Concepts and Design}, 5th~ed.\hskip 1em plus 0.5em minus 0.4em\relax
  USA: Addison-Wesley Publishing Company, 2011.

\bibitem{GeisbergerBroy2014}
E.~Geisberger and M.~Broy, Eds., \emph{Living in a networked world.
  {I}ntegrated research agenda {Cyber-Physical Systems (agendaCPS)}}, ser.
  acatech STUDY.\hskip 1em plus 0.5em minus 0.4em\relax Munich, Germany:
  Herbert Utz Verlag, 2014.

\bibitem{Schaetz2014}
B.~Schaetz, ``The role of models in engineering of cyber-physical systems –
  challenges and possibilities,'' in \emph{CPS20: CPS 20 years from now -
  visions and challenges}, ser. CPS Week, 2014.

\bibitem{Atzori+2010}
L.~Atzori, A.~Iera, and G.~Morabito, ``The internet of things: A survey,''
  \emph{Computer Networks}, vol.~54, no.~15, pp. 2787 -- 2805, 2010.

\bibitem{Harrand+2016}
N.~Harrand, F.~Fleurey, B.~Morin, and K.~E. Husa, ``{ThingML: A} language and
  code generation framework for heterogeneous targets,'' in \emph{Proceedings
  of the ACM/IEEE 19th International Conference on Model Driven Engineering
  Languages and Systems}, ser. MODELS '16, 2016.

\bibitem{KellyTolvanen2008}
S.~Kelly and J.-P. Tolvanen, \emph{Domain-Specific Modeling: Enabling Full Code
  Generation}, 1st~ed.\hskip 1em plus 0.5em minus 0.4em\relax Wiley-IEEE
  Computer Society Pr, 2008.

\bibitem{ThingML}
``{ThingML},'' \url{https://github.com/TelluIoT/ThingML}, accessed: 2020-04-29.

\bibitem{Moin+2018}
A.~Moin, S.~Rössler, and S.~Günnemann, ``{ThingML+}: Augmenting model-driven
  software engineering for the internet of things with machine learning,'' in
  \emph{Proceedings of the 2nd International Workshop on Model-Driven
  Engineering for the Internet of Things ({MDE4IoT})}, 2018.

\bibitem{Moin+2020}
A.~Moin, S.~Rössler, M.~Sayih, and S.~Günnemann, ``From things’ modeling
  language ({ThingML}) to things’ machine learning ({ThingML2}),'' in
  \emph{Proceedings of MODELS 2020 Satellite Events (Poster Companion /
  Extended Abstract), the ACM / IEEE 23rd International Conference on Model
  Driven Engineering Languages and Systems (MODELS)}, 2020.

\bibitem{OMGMDA2014}
\BIBentryALTinterwordspacing
``\BIBforeignlanguage{en}{{M}odel {D}riven {A}rchitecture ({MDA}), {MDA}
  {G}uide rev. 2.0},'' Object Management Group, Boston, MA, USA, Standard OMG
  Document ormsc/14-06-01, 2014. [Online]. Available:
  \url{https://www.omg.org/cgi-bin/doc?ormsc/14-06-01}
\BIBentrySTDinterwordspacing

\bibitem{Hall+2009}
M.~Hall, E.~Frank, G.~Holmes, B.~Pfahringer, P.~Reutemann, and I.~H. Witten,
  ``The weka data mining software: An update,'' \emph{SIGKDD Explor. Newsl.},
  vol.~11, no.~1, p. 10–18, Nov. 2009.

\bibitem{Bifet+2010}
A.~Bifet, G.~Holmes, R.~Kirkby, and B.~Pfahringer, ``Moa: Massive online
  analysis,'' \emph{J. Mach. Learn. Res.}, vol.~11, p. 1601–1604, Aug. 2010.

\bibitem{Pedregosa+2011}
F.~Pedregosa, G.~Varoquaux, A.~Gramfort, V.~Michel, B.~Thirion, O.~Grisel,
  M.~Blondel, P.~Prettenhofer, R.~Weiss, V.~Dubourg, J.~Vanderplas, A.~Passos,
  D.~Cournapeau, M.~Brucher, M.~Perrot, and E.~Duchesnay, ``Scikit-learn:
  Machine learning in {P}ython,'' \emph{Journal of Machine Learning Research},
  vol.~12, pp. 2825--2830, 2011.

\bibitem{Abadi+2015}
\BIBentryALTinterwordspacing
M.~Abadi, A.~Agarwal, P.~Barham, E.~Brevdo, Z.~Chen, C.~Citro, G.~S. Corrado,
  A.~Davis, J.~Dean, M.~Devin, S.~Ghemawat, I.~Goodfellow, A.~Harp, G.~Irving,
  M.~Isard, Y.~Jia, R.~Jozefowicz, L.~Kaiser, M.~Kudlur, J.~Levenberg,
  D.~Man\'{e}, R.~Monga, S.~Moore, D.~Murray, C.~Olah, M.~Schuster, J.~Shlens,
  B.~Steiner, I.~Sutskever, K.~Talwar, P.~Tucker, V.~Vanhoucke, V.~Vasudevan,
  F.~Vi\'{e}gas, O.~Vinyals, P.~Warden, M.~Wattenberg, M.~Wicke, Y.~Yu, and
  X.~Zheng, ``{TensorFlow}: Large-scale machine learning on heterogeneous
  systems,'' 2015, software available from tensorflow.org. [Online]. Available:
  \url{http://tensorflow.org/}
\BIBentrySTDinterwordspacing

\bibitem{Chollet+2015}
F.~Chollet \emph{et~al.}, ``Keras,'' \url{https://keras.io}, 2015.

\bibitem{Paszke+2017}
A.~Paszke, S.~Gross, S.~Chintala, G.~Chanan, E.~Yang, Z.~DeVito, Z.~Lin,
  A.~Desmaison, L.~Antiga, and A.~Lerer, ``Automatic differentiation in
  pytorch,'' in \emph{NIPS-W}, 2017.

\bibitem{HEADS}
``{HEADS},'' \url{https://cordis.europa.eu/project/id/611337}, accessed:
  2020-02-08.

\bibitem{Theano2016}
\BIBentryALTinterwordspacing
{Theano Development Team}, ``{Theano: A {Python} framework for fast computation
  of mathematical expressions},'' \emph{arXiv e-prints}, vol. abs/1605.02688,
  May 2016. [Online]. Available: \url{http://arxiv.org/abs/1605.02688}
\BIBentrySTDinterwordspacing

\bibitem{Mahout}
``{Apache Mahout},'' \url{https://mahout.apache.org}, accessed: 2020-09-12.

\bibitem{Meng+2016}
X.~Meng, J.~Bradley, B.~Yavuz, E.~Sparks, S.~Venkataraman, D.~Liu, J.~Freeman,
  D.~Tsai, M.~Amde, S.~Owen, D.~Xin, R.~Xin, M.~J. Franklin, R.~Zadeh,
  M.~Zaharia, and A.~Talwalkar, ``Mllib: Machine learning in apache spark,''
  \emph{J. Mach. Learn. Res.}, vol.~17, no.~1, p. 1235–1241, Jan. 2016.

\bibitem{Chen+2015}
T.~Chen, M.~Li, Y.~Li, M.~Lin, N.~Wang, M.~Wang, T.~Xiao, B.~Xu, C.~Zhang, and
  Z.~Zhang, ``Mxnet: A flexible and efficient machine learning library for
  heterogeneous distributed systems,'' 2015.

\bibitem{Berthold+2009}
\BIBentryALTinterwordspacing
M.~R. Berthold, N.~Cebron, F.~Dill, T.~R. Gabriel, T.~K\"{o}tter, T.~Meinl,
  P.~Ohl, K.~Thiel, and B.~Wiswedel, ``{KNIME - the Konstanz Information Miner:
  Version 2.0 and Beyond},'' \emph{SIGKDD Explor. Newsl.}, vol.~11, no.~1, pp.
  26--31, Nov. 2009. [Online]. Available:
  \url{http://doi.acm.org/10.1145/1656274.1656280}
\BIBentrySTDinterwordspacing

\bibitem{Bishop2013}
C.~M. Bishop, ``Model-based machine learning,'' \emph{Philosophical
  Transactions of the Royal Society A}, 2013.

\bibitem{InferNet}
T.~Minka, J.~Winn, J.~Guiver, Y.~Zaykov, D.~Fabian, and J.~Bronskill,
  ``{Infer.NET 0.3},'' 2018, microsoft Research Cambridge.
  http://dotnet.github.io/infer.

\bibitem{Breuker2014}
D.~Breuker, ``Towards model-driven engineering for big data analytics -- an
  exploratory analysis of domain-specific languages for machine learning,'' 01
  2014, pp. 758--767.

\bibitem{ML-Quadrat}
``{ML-Quadrat},'' \url{https://github.com/arminmoin/ML-Quadrat}, accessed:
  2020-09-12.

\bibitem{Newman1994}
W.~Newman, ``A preliminary analysis of the products of {HCI} research, using
  pro forma abstracts,'' in \emph{Proceedings of the SIGCHI Conference on Human
  Factors in Computing Systems}, ser. CHI '94.\hskip 1em plus 0.5em minus
  0.4em\relax New York, NY, USA: Association for Computing Machinery, 1994, p.
  278–284.

\bibitem{Basili+1994}
V.~Basili, G.~Caldiera, and H.~D. Rombach, ``{The Goal Question Metric
  Approach},'' \emph{Encyclopedia of Software Engineering - 2 Volume Set},
  1994.

\bibitem{KolterJohnson2011}
J.~Z. Kolter and M.~J. Johnson, ``{REDD}: A public data set for energy
  disaggregation research,'' in \emph{Proceedings of the {SustKDD} workshop on
  Data Mining Applications in Sustainability}, 2011.

\bibitem{KellyKnottenbelt2015}
\BIBentryALTinterwordspacing
J.~Kelly and W.~Knottenbelt, ``The {UK-DALE} dataset, domestic appliance-level
  electricity demand and whole-house demand from five {UK} homes,'' \emph{Sci
  Data 2}, March 2015. [Online]. Available:
  \url{https://doi.org/10.1038/sdata.2015.7}
\BIBentrySTDinterwordspacing

\bibitem{Murray+2017}
\BIBentryALTinterwordspacing
D.~Murray, L.~Stankovic, and V.~Stankovic, ``An electrical load measurements
  dataset of {United Kingdom} households from a two-year longitudinal study,''
  \emph{Sci Data 4}, January 2017. [Online]. Available:
  \url{https://doi.org/10.1038/sdata.2016.122}
\BIBentrySTDinterwordspacing

\bibitem{Selic2007}
B.~Selic, ``A systematic approach to domain-specific language design using
  uml,'' 05 2007, pp. 2--9.

\bibitem{Lagarde+2007}
\BIBentryALTinterwordspacing
F.~Lagarde, H.~Espinoza, F.~Terrier, and S.~G\'{e}rard, ``Improving uml profile
  design practices by leveraging conceptual domain models,'' in
  \emph{Proceedings of the Twenty-Second IEEE/ACM International Conference on
  Automated Software Engineering}, ser. ASE '07.\hskip 1em plus 0.5em minus
  0.4em\relax New York, NY, USA: Association for Computing Machinery, 2007, p.
  445–448. [Online]. Available: \url{https://doi.org/10.1145/1321631.1321705}
\BIBentrySTDinterwordspacing

\end{thebibliography}
\bibliographystyle{IEEEtran}

\end{document}